\documentclass[journal=aelccp,manuscript=article]{achemso}
\usepackage{color}
\usepackage[version=3]{mhchem} % Formula subscripts using \ce{}
\usepackage{pdfpages}

\author{Zijian Hong}
\affiliation{Department of Mechanical Engineering, Carnegie Mellon University, Pittsburgh, Pennsylvania 15213, USA}
\author{Venkatasubramanian Viswanathan}
\affiliation{Department of Mechanical Engineering, Carnegie Mellon University, Pittsburgh, Pennsylvania 15213, USA}
\email{venkvis@cmu.edu}

\title{Phase-field simulations of lithium dendrite growth with open-source software}

\begin{document}

\begin{tocentry}
\begin{center}
\includegraphics[width=9cm,height=3.5cm,keepaspectratio]{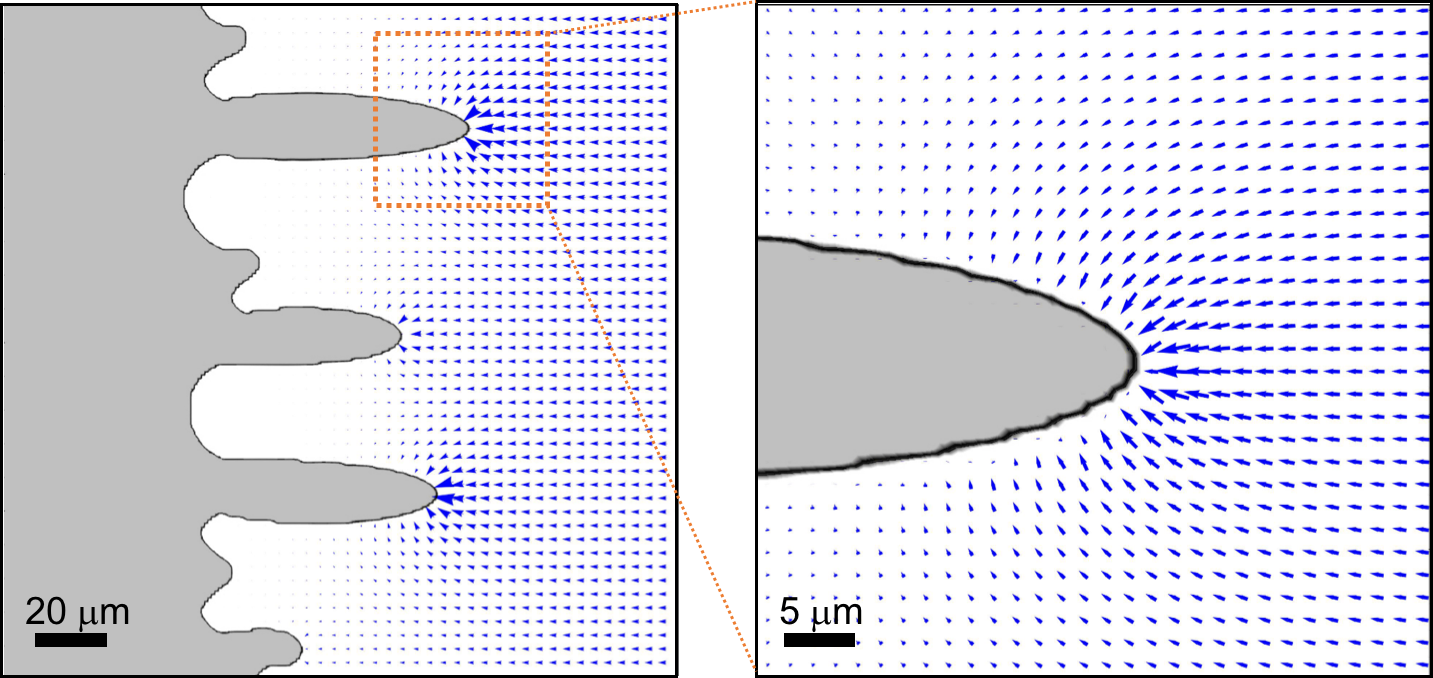}
\end{center}
\end{tocentry}

\begin{abstract}
Dendrite growth is a long-standing challenge that has limited the applications of rechargeable lithium metal electrodes. Here, we have developed a grand potential based nonlinear phase-field model to study the electrodeposition of lithium as relevant for a lithium metal anode, using open-source software package MOOSE. The dynamic morphological evolution under large/small overpotential is studied in 2-dimensions, revealing important dendrite growth/stable deposition patterns. The corresponding temporal-spatial distributions of ion concentration, overpotential and driving force are studied, which demonstrate an intimate, dynamic competition between ion transport and electrochemical reactions, resulting in vastly different growth patterns. Given the importance of morphological evolution for lithium metal electrodes, wide-spread applications of phase-field models have been limited in part due to in-house or proprietary software.  In order to spur growth of this field, we utilize an open-source software package and make all files available to enable future studies to study the many unsolved aspects related to morphology evolution for lithium metal electrodes.  
\end{abstract}

\newpage
There is a need for increasing the energy density of batteries for enabling a complete electrification of transportation.\cite{Sripad01012017,doi:10.1021/acsenergylett.7b00432,doi:10.1021/acsenergylett.7b01022} Lithium metal anodes represent one of the most promising near-term solutions enabling increased energy density. \cite{kerman2017practical}  The main challenge with lithium metal anodes stems from dendrite growth during electrodeposition, which leads to deposits that could penetrate through the separator, resulting in safety issues and coloumbic inefficiency.\cite{AURBACH2002405,doi:10.1021/acs.chemrev.7b00115,C3EE40795K,Lin2017}  A comprehensive understanding of the morphology evolution during the electrodeposition process is of paramount importance for the design of next-generation batteries with Lithium metal anode.

Most of the previous theoretical investigations of dendrite formation during electrodeposition is based on linear stability analysis, originally pioneered by Mullins and Sekerka \cite{MullinsSekerka63,MullinsSekerka64}, where the thermodynamic stability of a small perturbation is studied in terms of a change in the chemical potential\cite{monroe2003-dendrite,Monroe2004Effect,Monroe2005Impact,Tikekar2016,ahmad2017stability}. This simple approach has revealed the impact of some key physical parameters (elastic properties, molar volume, etc.\cite{Tikekar2016,ahmad2017stability,PhysRevMaterials.1.055403}).  However, the linear stability analysis approach is limited to the understanding of dendrite initiation and does not take into account effects beyond nucleation, dimensionality of growth and dynamics of the growth process.  In particular, the key features related to morphology evolution in three-dimensions are missed within this approach. 

Phase-field modeling is a mesoscale simulation tool that enable the 
quantitative understanding of phase transitions, phase transformations and microstructure evolutions. \cite{doi:10.1146/annurev.matsci.32.112001.132041,doi:10.1146/annurev.matsci.32.101901.155803}  To date, several phase-field models have been built to model the dynamic evolution of dendrite growth during the electrodeposition process, incorporating the nonlinear electrochemical reaction dynamics \cite{CHEN2015376,PhysRevE.92.011301,Liang2014,Liang2012,ELY2014581,doi:10.1021/ar300145c}. However, these studies typically utilize custom-built software or proprietary code which could hinder the wide-spread applications of phase-field models, while also makes it extremely difficult for follow-on works and reproducible science.\cite{Kitchin:2015aa}

Herein, we developed a hybrid grand potential based nonlinear phase field model \cite{PhysRevE.92.011301,CHEN2015376,PhysRevE.84.031601}, and numerically solving the multiphysics coupled equations using the fully \textit{open source} MOOSE (Multiphysics Object-Oriented Simulation Environment) framework \cite{GASTON20091768}.  This enables a quantitative and comprehensive understanding of the morphological evolution during the electrodeposition process. 
The well tested and documented MOOSE package allows for fast, parallel and customized numerical modeling, which is easy to follow and simple to implement and validate. Most importantly, it has an already built-in, well-benchmarked phase-field module that could be utilized for further implementations \cite{TONKS201220,YURKIV2018609}. In this work, the dynamic morphological evolution for lithium deposition is studied, where both stable deposition and dendrite growth are captured at low and high applied overpotentials, respectively. The corresponding spatial-temporal distributions of concentration, over-potential and driving force are studied in details, which reveals an intimate, dynamic competition between lithium-ion transport (from both ion diffusion and electrical migration) and electrochemical reaction, resulting in complete different growth patterns. A concept of ``compositionally graded solid electrolyte'' is proposed, which could provide an approach to suppress dendrite initiation based on the insights from this study.   We believe that this study will spur on the application of phase-field models that can handle more complicated problems incorporates other effects (elastic effect, solid electrolyte interfaces, etc.).

\textit{Phase field model.} The phase field variable $\xi$ is used as the non-conserved order parameter which is defined such that $\xi=1$  and $\xi=0$ for the pure electrode and electrolyte phases, respectively. Following the derivations by Chen et al.\cite{CHEN2015376}, the temporal evolution of the order parameter $\xi$ can be expressed as:
\begin{equation}
\frac{\partial\xi}{\partial t}=-L_\sigma(g'(\xi)-k\nabla^2\xi)-L_\eta h'(\xi) \Bigg\{\exp\Big[\frac{(1-\alpha)nF\eta_{\alpha}}{RT}\Big]-\frac{c_{Li^+}}{c_0}\exp\Big[\frac{-\alpha nF\eta_{\alpha}}{RT}\Big]\Bigg\}
.
\end{equation}
where $L_\sigma$ and $L_\eta$ are the interfacial mobility and electrochemical reaction kinetic coefficient, respectively. $g(\xi)$  is the double well function, expressed by $W\xi^2(1-\xi)^2$. $W$ is related to the switching barrier \cite{CHEN2015376}. $k, \alpha, n, F, R, t$ and $T$ are the gradient coefficient, charge transfer coefficient, number of electrons transfered, Faraday constant, gas constant, evolution time and temperature, respectively. The switching barrier and gradient coefficient are further related to the surface tension $\gamma$ and interfacial thickness $\delta$, i.e. $W=\frac{3\gamma}{\delta}$ and $k=6\gamma \delta$ \cite{ELY2014581,doi:10.1146/annurev.matsci.32.101901.155803}. Whereas $h(\xi)=\xi^3(6\xi^2-15\xi+10)$ is the interpolation function \cite{CHEN2015376}. $\eta_{\alpha}$ is the activation overpotential $\eta_{\alpha}=\phi-E^\theta$,$E^\theta$ is the standard equilibrium half cell potential, $\phi$ is the applied overpotential. $c_{Li^+}$ and $c_0$ are the local and initial lithium-ion molar ratio.
In a two phase model, the local lithium ion molar-fraction can be written as \cite{PhysRevE.92.011301}:
\begin{equation}
c_{Li^+}=c^l (1-h(\xi))=\frac{exp[\frac{(\mu-\epsilon^l)}{RT}]}{1+exp[\frac{(\mu-\epsilon^l)}{RT}]} (1-h(\xi))
.
\end{equation}
where $c^l$ is the molar fraction of lithium in the electrolyte phase, $\mu$ is the chemical potential of lithium. $\epsilon^l=\mu^{0l}-\mu^{0N}$ is the differences in the chemical potential of lithium and neutral components on the electrolyte phase at initial equilibrium state.

The chemical potential can be obtained by solving the modified diffusion equation \cite{PhysRevE.92.011301} (detailed derivation is given in Supplementary Information):
\begin{equation}
\frac{\partial \mu}{\partial t}=\frac{1}{\chi}[\nabla.\frac{Dc_{Li^+}}{RT}(\nabla\mu+nF\nabla\phi)-\frac{\partial h(\xi)}{\partial t} (c^s \frac{C_m^s}{C_m^l}-c^l)]
\end{equation}
where $C_m^s$ and  $C_m^l$ are the site density of the electrode and electrolyte phases (inverse of molar volume), respectively. $c^s$ is the molar fraction of lithium in the electrode phase, while $\chi=\frac{\partial{c^l}}{\partial \mu}[1-h(\xi)]+\frac{\partial{c^s}}{\partial \mu}h(\xi)\frac{C_m^s}{C_m^l}$ . 

The spatial distribution of the electrical overpotential $\phi$ can be obtained by solving the conduction equation:
\begin{equation}
\nabla\sigma\nabla\phi=nFC_m^s\frac{\partial\xi}{\partial t}
.
\end{equation}
The effective conductivity $\sigma$ is related to the conductivity of the electrode phase $\sigma^s$ and electrolyte phase $\sigma^l$ , i.e., $\sigma=\sigma^s h(\xi)+\sigma^l [1-h(\xi)]$

\textit{Numerical simulations.} Equations 1-4 are solved using the open source MOOSE framework. A two-dimensional mesh of 200$\times$200 is used, with each grid representing 1 $\mu$m. The simulation temperature is 300 K. The Preconditioned Jacobian Free Newton Krylov (PJFNK) method is set as the solve type, with a bdf2 scheme and Single Matrix Preconditioning (SMP). A timestep interval (dt) of 0.02 s is set, with a maximum simulation time of 400 s. The key simulation parameters before and after normalization are listed in Supplementary table 1. The normalization constants for length, time and energy scales are set as 1 $\mu$m, 1 s and $2.5 \times 10^6 $J/m$^3$. The standard equilibrium half cell potential is set to be 0 for the sake of convenience. A Langevin noise level of 0.04 is added (corresponding to 0.056\% change in magnitude) to the order parameter $\xi$ to stimulate the thermal fluctuation in the system. The Neumann boundary condition is applied at the $X$-dimension, such that the chemical potential is fixed to be zero at both sides (indicating a constant Lithium molar ratio at the two ends), while the order parameter is set as 1 (electrode phase) and 0 (electrolyte phase) at the left and right sides, respectively. The electric overpotential is  fixed to the applied overpotential at the electrode side and zero at the electrolyte side to give a constant activation overpotential across the electrode/electrolyte interface. The no flux boundary condition is applied along the $Y$-dimension for all primary variables ($\xi, \mu$ and $\phi$). A set of applied electric overpotential ranging from -0.3 V to -0.5 V is studied.

\begin{figure}[!htb] 
  \centering 
  \includegraphics[width=1.0\linewidth]{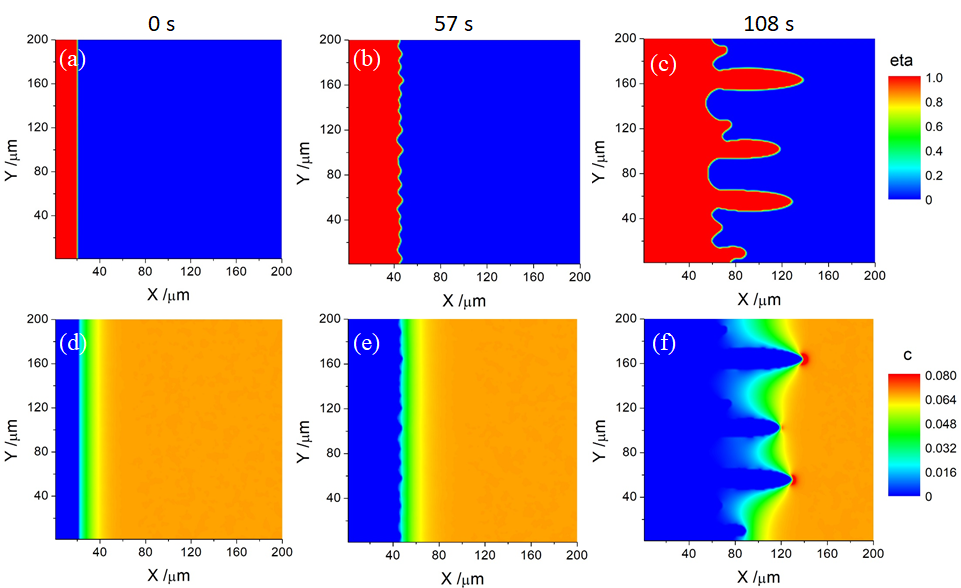}
 \caption{Dendrite growth and concentration evolution with applied overpotential of -0.45 V. (a)-(c) Morphology evolution after Lithium electrodeposition of 0 s, 57 s and 108 s. (d)-(f) Corresponding evolution of Lithium-ion molar ratio.} 
 \label{fig:example1} 
\end{figure}

\textit{Results.} The lithium growth kinetics under an applied overpotential of -0.45 V is shown in Figure~\ref{fig:example1}, assuming a Li$^+$ concentration of 1 M in a standard carbonate-based electrolyte. The initial structure is shown in Figure~\ref{fig:example1}(a), an electrode with a thickness of 20 $\mu$m is separated with the electrolyte by a smooth interface. No initial nuclei are added, the dendrite initiation and growth is a purely intrinsic process in this study. This differs from several previous studies where the simulation results might be sensitive to the preset artificial nuclei size, shape and orientation \cite{CHEN2015376,YURKIV2018609,ELY2014581}. After an electrodeposition process of 57 s (Figure~\ref{fig:example1}b), the interface moves towards the electrolyte by $\sim$20 $\mu$m. The interfacial instability is discovered, with a modulation of surface roughness for a vertical period of $\sim$10 $\mu$m. Further growth of these nuclei leads to the formation of long needle-like dendrites (Figure~\ref{fig:example1}c) with a length of $\sim$50 $\mu$m, while the diameter of the dendrites only increases by $\sim$10 $\mu$m, indicating a quasi-directional growth. A short video of the dendrite growth process under this condition is given in Supplementary video 1. It is worthwhile noting that although $\sim$20 nuclei is formed initially, only a few of them can grow into dendrites.
The corresponding evolution of Li$^+$ ion molar fraction is shown in Figures~\ref{fig:example1}(d)-(f). It is revealed that a broad interfacial region (shown in green) is formed to separate the electrode (with zero Li$^+$ molar ratio) and the electrolyte (with a homogeneous Li$^+$ molar ratio of $\sim$0.067). With the formation of small nuclei at the interface (Figure~\ref{fig:example1}e), a small fluctuation of ion concentration is also formed accordingly where the tip region has a slightly higher concentration than the valley region. The further growth of dendrites leads to a higher Li$^+$ concentration surrounding the dendrite tips, with a lower ion concentration formed in the valley regions. The enrichment of Li$^+$ molar ratio in the vicinity of the  dendrite tips is due to the shortened diffusion path compared to the valley regions.

\begin{figure}
  \centering 
  \includegraphics[width=0.88\linewidth]{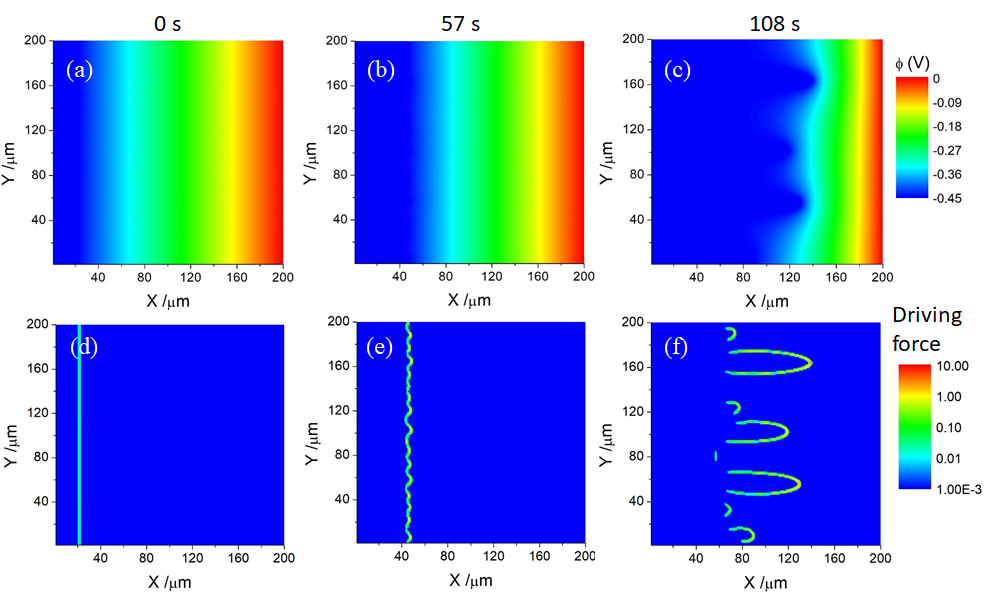} \includegraphics[width=0.88\linewidth]{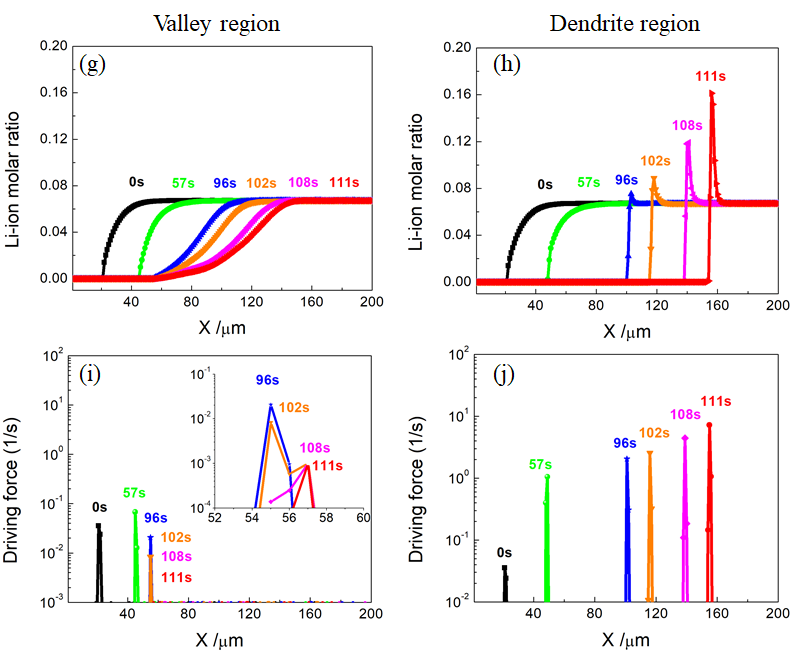}
 \caption{Spatial/temporal  distribution of electric potential and driving force. (a)-(c) Evolution of electric potential at 0 s, 57 s and 108 s. (d)-(f) The corresponding spatial distribution of the driving force. (g)-(h) The line plots of Li-ion molar ratio as a function of time in the Valley and Dendrite regions. (i) and (j) The corresponding Line plots of the driving forces in the two regions.} 
 \label{fig:example2} 
\end{figure}

In order to take a deeper understanding of the dendrite growth mechanism, the spatial-temporal distribution of electric potential and driving force is plotted in Figure~\ref{fig:example2}. As shown in Figure~\ref{fig:example2}(a), the electric potential is nearly constant in the electrode region, followed by a gradual, linear increase from the electrode/electrolyte interface to the  electrolyte region (see Figure S2d). After 57 s (Figure~\ref{fig:example2}b), as the interface is moving towards the electrolyte region, the gradient of electric potential (electric field) increases. The electric field is still more or less homogeneous inside the electrolyte region. With the growth of dendrite (Figure~\ref{fig:example2}c), a lower electric potential is found near the tip as compared to the surrounding valley regions, thus, we expect not only a large electric field along the $X$-dimension, but also an electric field towards the dendrite tip along the $Y$-dimension as well (shown in Figure S3).   This could further facilitate Li$^+$ transport from both the source and the surrounding valley regions to the dendrite tips, leading to an ``entrainment'' phenomenon. The driving force (defined as $L_\eta h'(\xi) \Bigg\{\exp\Big[\frac{(1-\alpha)nF\eta_{\alpha}}{RT}\Big]-\frac{c_{Li^+}}{c_0}\exp\Big[\frac{-\alpha nF\eta_{\alpha}}{RT}\Big]\Bigg\}$) for the electrochemical reaction is directly related to the interface growth velocity $\frac{\partial\xi}{\partial t}$. As depicted in Figure 2(d), there's a large driving force at the electrode/electrolyte interface, which can be understood since the electrochemical reaction only occurs at the interface where $h'(\xi)$ is non-zero. After 57 s (Figure 2e), with the continuous electrochemical deposition, the driving force at the interface increases, while forming a zig-zag like pattern, mimicking the modulation of the solid/liquid interface. With the growth of the dendrites after 108 s (Figure 2f),it is clearly shown that the driving force at the dendrite tips are orders of magnitude larger than the interface of the valley regions. Whereas the longer the dendrite, the larger the driving force, which further facilitates growth of the longer dendrites. Figure 2(g) and (h) show the line plots for the time evolution of Li-ion molar fraction along $X$-direction cutting through the valley and dendrite regions, respectively. It can be seen that up until 57 s where the dendrites are yet to grow, the concentration profile in the two regions are very similar, the interface region has a lower Li$^+$ ion molar fraction as compared to the bulk. As the dendrites grow longer (e.g., 96 s), we see an enrichment/depletion of ion concentration at the dendrite/valley regions, resulting in a widely broadened interface for the valley region which can hardly move thereafter. The line plots of the driving force for the two regions are shown in figure 2(i) and (j). Spikes are found at the interfaces in accordance to the  concentration profiles. After 57 s, as the interface is moving towards the electrolyte, the electric field increases, which increases the driving force. Eventually, the growth of dendrites will lead to a huge increase in the driving force which can be partly attributed to the increase of ion concentration surrounding the dendrite tip; whereas the valley region gets an order of magnitude suppression in the driving force, resulting from the depletion of Li$^+$ ion.

\begin{figure} 
  \centering 
  \includegraphics[width=1.0\linewidth]{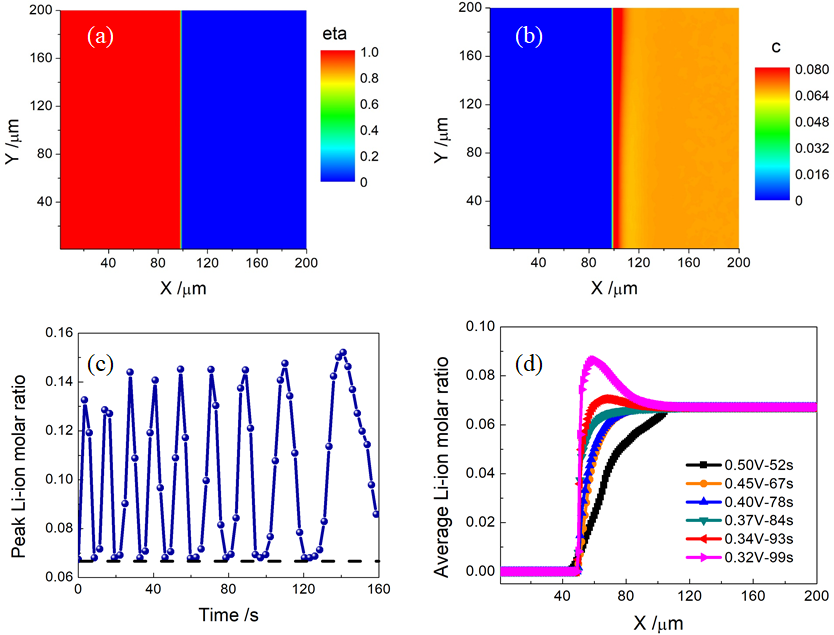}
 \caption{Influence of the applied overpotential. (a) Stable Lithium deposition after 200 s with an applied overpotential of -0.32 V. (b) The corresponding spatial distribution of ion concentration. (c) The temporal evolution of the maximum Li-ion concentration. The dashed line shows the bulk concentration. (d) Average Li-ion concentration profile at different applied overpotential after passing the same total current. } 
 \label{fig:example3} 
\end{figure}

The morphology evolution at low applied overpotential, e.g., -0.32 V is further studied to make a detailed comparison (Figure~\ref{fig:example3}). A stable electrochemical deposition is observed with no obvious surface modulation (Figure 3a). Interestingly, near the electrode/electrolyte interface, an enrichment in the Li$^+$ ion is discovered (Figure 3b). This can be understood since the deposition velocity is greatly reduced (exponentially) at lower overpotential, while the ion transport only depends on the gradient of the potential distribution which roughly decreases linearly with decreasing applied overpotential. The temporal evolution of the maximum Li$^+$ concentration near the interface is plotted in Figure 3(c). An oscillation of the maximum Li$^+$ ion molar ratio is captured, indicating a dynamic competition between the ion transport and electrochemical reaction. The line plot of the average Li$^+$ concentration profile after passing the same total current is plotted for different overpotentials (Figure 3d), a gradual shift from surface concentration enrichment to surface concentration depletion is observed from lower to higher overpotentials (0.32 V-0.5 V). And the threshold overpotential for dendrite growth occurs exactly at the transition point from interface concentration enrichment to depletion ($\sim$0.37 V). 

\textit{Discussions.} The schematics of the two scenarios are given below for a discussion. At lower applied overpotential, initially, the reaction rate is much smaller than the ion transport rate (including diffusion and migration), resulting in an accumulation of ion concentration at the interface, which would in turn boost the reaction rate (since it is directly proportional to the ion concentration) while decreasing the transport rate (due to a large negative concentration gradient). With the increase of reaction rate and decrease of transport rate, the transport rate will be surpassed by the reaction rate , giving rise to a decrease in surface ion concentration. And as it happens, the decrease of surface ion concentration would lead to an increase in ionic transport and decrease of electrochemical reaction. Thus, an iterative feedback system is achieved to guarantee the stable electrodeposition. Importantly, in this case the maximum surface ion concentration is always higher than the concentration of the bulk electrolyte. 

\begin{figure} 
  \centering 
  \includegraphics[width=1.0
  \linewidth]{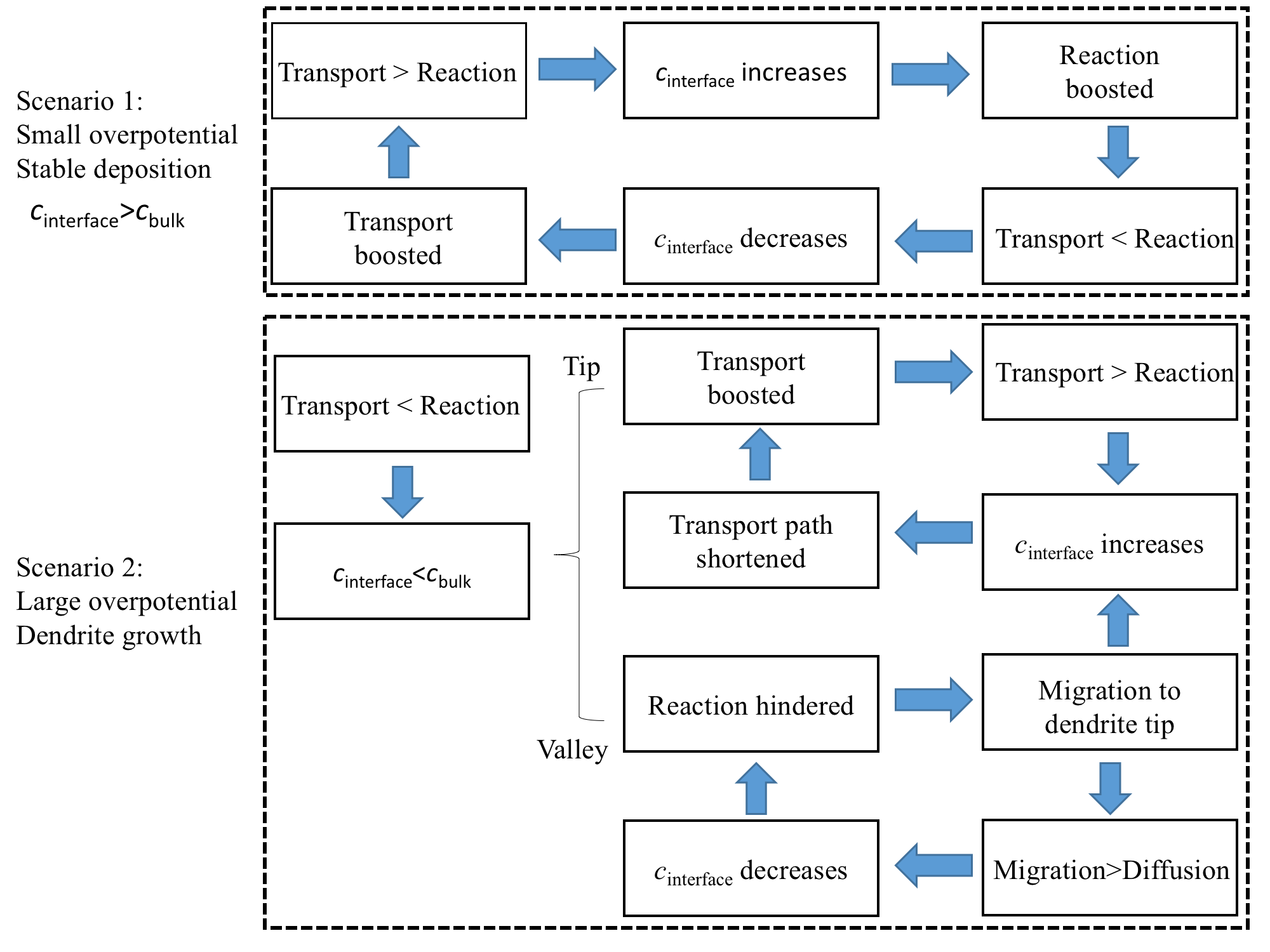}
 \caption{Schematics of the dynamic evolution scenarios at small and large overpotentials (low and high currents).} 
 \label{fig:example4} 
\end{figure}

While under higher overpotential, initially, the electrochemical reaction is faster than the ion transport, resulting in a depletion of ions at the interface. The formation of peaks (dendrites) due to the inhomogeneity of the system will boost the transport of Li$^+$, due to not only a larger concentration gradient, but also a higher migration rate from both the source and the surrounding valley regions. This results in a higher transport rate than the reaction rate, leading to a further increase in the surface ion concentration. The increase in surface ion concentration in return boosts the electrochemical reaction and interfacial growth velocity which further shortens the diffusion length. In other surrounding valley regions, the ions are ``sucked'' by the peak region, the ion concentration keeps dropping which hinders the electrochemical reaction.   This effect is purely related to the dimensionality and cannot be captured within 1-dimensional models.  
The key feature that distinguishes the two scenarios is that the average ion concentration at the interface is always larger than the bulk electrolyte for the lower overpotential case, while vice versa for the higher overpotential. This indicates that even when a surface instability is formed in the first scenario, it will quickly be recovered due to a lower reaction rate with lower concentration at the tip of the surface modulation.

Based on the above understanding, we proposed a new concept ``Compositionally graded solid electrolyte'' which could potentially suppress the dendrite initiation (given in Supplementary Figure S6). This electrolyte is designed such that a higher Li-ion concentration at the Li-metal anode side and a lower Li-ion concentration at the cathode side. This can be achieved by tuning the polymer/ceramics ratio, or by varying the vacancy/doping concentrations in the solid solutions. Under this design, the initiation of dendrite would be hindered at a reasonable current density since the tips (dendrite) would have a lower concentration than the valley regions, which would decrease the electrochemical reaction rate of the dendrite.   

In conclusion, we have developed a hybrid grand potential based nonlinear phase-field model for electrodeposition using fully \textit{open source} MOOSE package. The dynamic morphological evolution during the lithium electrodeposition process under large and small overpotential is studied, revealing dendrite growth/stable deposition under the two conditions. The concentration, overpotential, driving force distribution is further studied for both cases, showing the critical role of the intimate competitions between ion transport and electrochemical reactions for the formation of dendrites. We further proposed a ``compositionally graded solid electrolyte'' that could potentially lead to the suppression of dendrites.  We hope that open-sourcing phase-field models will benefit the community as a whole, and stimulate more theoretical/experimental follow-up studies.

%\textit{Outlook -- many more things should be addressed. self-formed SEI, solid electrolyte, anisotropy.  We hope that open-sourcing this will enable many others to explore these avenues.} 

\begin{acknowledgement}

Z.H. and V.V. gratefully acknowledge support from the U.S. Department of Energy, Energy Efficiency and Renewable Energy Vehicle Technologies Office under Award No. DE-EE0007810. 
\end{acknowledgement}

\begin{suppinfo}
All input files for MOOSE simulations are made freely available. Contents in the supporting information: Derivation of the modified diffusion equation; Parameters used in the simulations; Nonlinear dendrite growth kinetics; Branching growth kinetics under high overpotential; Design of ``compositionally graded solid electrolyte''; Supplementary videos for the morphology and concentration evolution under applied overpotential of -0.45 V and -0.32 V.
\end{suppinfo}

\bibliography{refs}

\includepdf[pages={1-}]{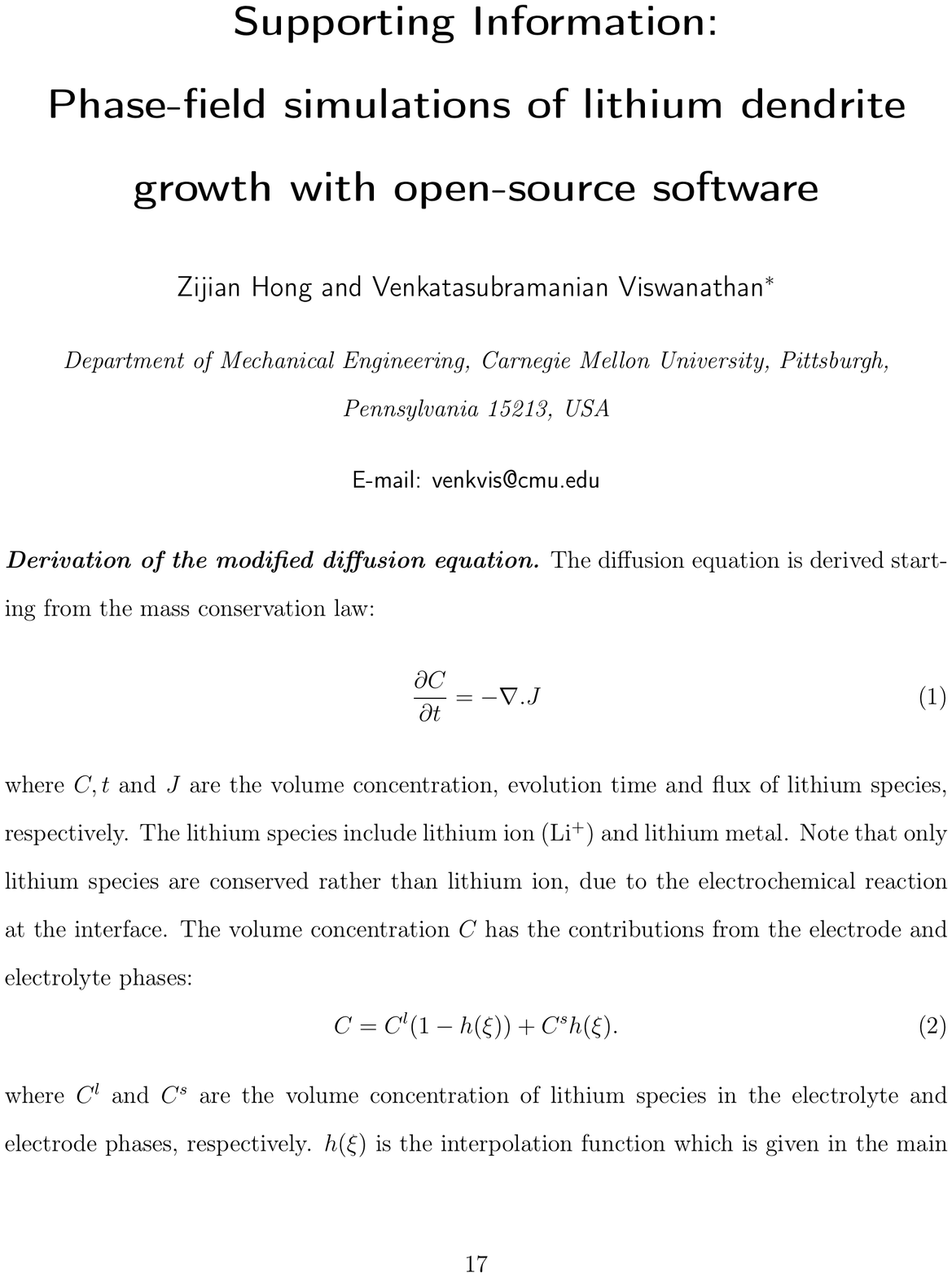}

\end{document}